\documentclass[prb,twocolumn,showpacs,epsfig,epsf]{revtex4}
\usepackage{amsmath,amsthm,amsfonts,amscd,mathrsfs,pifont,bm}
\usepackage{eucal,graphicx}
\renewcommand{\atop}[2]{\genfrac{}{}{0pt}{}{#1}{#2}}

\begin{document}

\title  {Statistics of reflection eigenvalues in chaotic cavities with non-ideal leads}

\author {Pedro Vidal and Eugene Kanzieper}

\affiliation
       {
       Department of Applied Mathematics, H.I.T.---Holon Institute of
       Technology,
       Holon 58102, Israel
       }
\date   {November 23, 2011}

\begin  {abstract}
The scattering matrix approach is employed to determine a joint probability density function of reflection eigenvalues for chaotic cavities coupled to the outside world through both ballistic and tunnel point contacts. Derived under assumption of broken time-reversal symmetry, this result is further utilised to (i) calculate the density and correlation functions of reflection eigenvalues, and (ii) analyse fluctuations properties of the Landauer conductance for the illustrative example of asymmetric chaotic cavity. Further extensions of the theory are pinpointed.
\end{abstract}

\pacs   {73.23.--b, 05.45.Mt, 02.30.Ik}
\maketitle
\newpage
{\it Introduction.}---At low temperatures and voltages, a phase coherent charge transfer through quantum chaotic cavities is known to exhibit a high degree of {\it statistical universality} \cite{B-1997,A-2000}. Even though the transport through an individual chaotic structure is highly sensitive to its microscopic parameters, the universal statistical laws emerge upon appropriate ensemble or energy averaging procedure. The latter efficiently washes out all system-specific features provided a charge carrier has stayed in a cavity long enough to experience diffraction \cite{AL-1996,RS-2002}. Quantitatively, this requires the average electron
dwell time $\tau_{\rm D}$ to be in excess of the Ehrenfest time $\tau_{\rm E}$ that defines the time scale where quantum effects set in.

In the extreme limit $\tau_{\rm D} \gg \tau_{\rm E}$, the statistics of charge transfer is shaped by the {\it underlying symmetries} \cite{B-1997} of a scattering system (such as the absence or presence of time-reversal, spin-rotational, and/or particle-hole symmetries). For this reason, a stochastic approach \cite{LW-1991} based on
the random matrix theory \cite{M-2004} (RMT) description \cite{BGS-1984} of electron dynamics in a cavity is naturally expected to constitute an efficient framework for nonperturbative studies of the universal transport regime. Indeed, a stunning progress was achieved in the RMT applications to the transport problems over the last two decades. Yet, intensive research in the field\cite{BFS-2011} left unanswered many basic-level questions. One of them, regarding the statistics of transmission/reflection eigenvalues in chaotic cavities coupled to the leads through the {\it point contacts with tunnel barriers}, will be a focus of this Letter.

Supported by the supersymmery field theoretic technique \cite{E-1997} as well as by recent semiclassical studies \cite{RS-2002}, the RMT approach to quantum transport starts with the Heidelberg formula for the scattering matrix \cite{MW-1963}
\begin{equation}
\label{sm}
    {\bm {\mathcal S}}(\varepsilon_F) = \openone_N - 2i \pi  {\bm {\mathcal W}}^\dagger
    ({\varepsilon}_F\,\openone_M - {\bm{\mathcal H}} + i \pi {\bm{\mathcal W}}
    {\bm {\mathcal W}}^\dagger)^{-1} {\bm{\mathcal W}}
\end{equation}
of the total system comprised by the cavity and the leads. Here, an $M \times M$ random matrix ${\bm {\mathcal H}}$ (of proper symmetry, $M\rightarrow \infty$) models a single electron Hamiltonian whilst
an $M\times N$ deterministic matrix ${\bm {\mathcal W}}$ describes the coupling of electron
states with the Fermi energy ${\varepsilon_F}$ in the cavity to
those in the leads; $N=n_{\rm L} + n_{\rm R}$ is the total number of
propagating modes (channels) in the left ($n_{\rm L}$) and right ($n_{\rm R}$) leads. Equation (\ref{sm}) refers to chaotic cavities with sufficiently large capacitance (small charging energy) when the electron-electron interaction can be disregarded \cite{B-1997}. Throughout the paper, only such cavities are considered.

Landauer's insight \cite{LFLB-1957} that electronic conduction in solids can be thought of as a scattering problem makes the $N\times N$ scattering matrix ${\bm {\mathcal S}}(\varepsilon_F)$ a central player in statistical analysis of various transport observables. In the physically motivated $M\rightarrow \infty$ scaling limit, its distribution, dictated solely by the symmetries of the random matrix $\bm{\mathcal H}$, is well studied for both normal \cite{B-1995} and normal-superconducting \cite{B-2009} chaotic systems. In the former case, the distribution of
${\bm {\mathcal S}}(\varepsilon_F)$ is described by the Poisson kernel \cite{H-1963,B-1995}
\begin{equation}
\label{pk}
    P_\beta({\bm {\mathcal S}}) \propto \big[
    {\rm det}(
        \openone_N -  \bar{\bm {\mathcal S}} {\bm {\mathcal S}}^\dagger)\,
        {\rm det}(
        \openone_N -   {\bm {\mathcal S}} \bar{\bm {\mathcal S}}^\dagger)
    \big]^{ \beta/2 -1  -  \beta N/2}.
\end{equation}
Here, $\beta$ is the Dyson index \cite{M-2004} accommodating system
symmetries: ${\bm {\mathcal S}}(\varepsilon_F)$ is unitary symmetric for $\beta=1$, unitary for $\beta=2$, and unitary self-dual for $\beta=4$. All relevant microscopic details of the scattering system are encoded into a single average scattering matrix
\begin{eqnarray}
    \bar{\bm {\mathcal S}} = (M\Delta \openone_N - \pi^2 {\bm {\mathcal W}}^\dagger {\bm {\mathcal W}})
    \,(M\Delta \openone_N + \pi^2 {\bm {\mathcal W}}^\dagger {\bm {\mathcal W}})^{-1},
\end{eqnarray}
where $\Delta$ denotes the mean level spacing at the Fermi level in the limit $M\rightarrow \infty$. The $N$ eigenvalues $\hat{\bm \gamma} = {\rm diag} (\{\sqrt{1-\Gamma_j}\})$ of $\bar{\bm {\mathcal S}}$ characterise\cite{B-1997} couplings between the cavity and the leads in terms of tunnel probabilities $\Gamma_j$ of the $j$-th electron mode in the leads. The celebrated result Eq.~(\ref{pk}), that can be viewed as a generalisation
of the three Dyson circular ensembles \cite{M-2004}, was alternatively derived through a phenomenological information-theoretic approach reviewed in Ref. \cite{MB-1999}.

Unfortunately, statistical information accommodated in the Poisson kernel is too detailed to make a nonperturbative description of transport observables {\it operational}. It turns out, however, that in case of conserving charge transfer through normal chaotic structures, it is suffice to know a probability measure associated with a set ${\bm T}$ of non-zero transmission eigenvalues $\{T_j \in (0,1)\}$; these are the eigenvalues of the Wishart-type matrix ${\bm t}{\bm t}^\dagger$, where ${\bm t}$ is the transmission sub-block of the scattering matrix
\begin{eqnarray}\label{s-block}
    \bm{{\mathcal S}} = \left(
                      \begin{array}{cc}
                 {\bm r}_{n_{\rm L}\times n_{\rm L}} & {\bm t}_{n_{\rm L}\times n_{\rm R}} \\
                 {\bm t^\prime}_{n_{\rm R}\times n_{\rm L}} & {\bm r^\prime}_{n_{\rm R}\times n_{\rm R}} \\
               \end{array}
                   \right).
\end{eqnarray}
Owing to this observation, the joint probability density function $P_\beta({\bm T})$ emerges as the object of primary interest in the RMT theories of quantum transport.

Surprisingly, our knowledge of the probability measure $P_\beta({\bm T})$ induced by the Poisson kernel [Eq.~(\ref{pk})] is very limited, being restricted to chaotic cavities coupled to external reservoirs via {\it ballistic point contacts}\cite{vHB-1996} (``ideal leads''). In this, mathematically simplest case, the unity tunnel probabilities $\Gamma_j=1$ make
the average scattering matrix $\bar{\bm {\mathcal S}}$ vanish, giving rise to the uniformly distributed\cite{BS-1990} scattering matrices which otherwise maintain a proper symmetry\cite{M-2004}. In the RMT language, this implies that scattering matrices belong to one of the three Dyson circular ensembles\cite{D-1962}.

As was first shown by Baranger and Mello\cite{BM-1994}, and by Jalabert, Pichard, and Beenakker\cite{JPB-1994}, the uniformity of scattering matrix distribution induces a nontrivial joint probability density function of transmission eigenvalues $\{T_j\}$ of the form\cite{F-2006}
\begin{equation}
\label{PnT}
    P_0^{(\beta)}({\bm T}) \propto  \,|\Delta_n^\beta({\bm T})| \prod_{j=1}^n
    T_j^{\beta/2-1 + \beta \nu /2}.
\end{equation}
Here, $\nu= |n_{\rm L}-n_{\rm R}|$ is the asymmetry parameter, $n = {\rm min}(n_{\rm L},n_{\rm R})$ is the number of non-zero eigenvalues of the matrix ${\bm t} {\bm t}^\dagger$, whilst $\Delta_n({\bm T})$ is the Vandermonde
determinant $\Delta_n({\bm T})=\prod_{j<k} (T_k-T_j)$. Equation (5) is one of the cornerstones of the RMT approach to quantum transport.

{\it From ballistic to tunnel point contacts.}---The restricted validity of Eq.~(\ref{PnT}), that holds true for chaotic cavities with {\it ideal leads}, is hardly tolerable both theoretically (an important piece of the transport theory is missing \cite{R-previous}) and experimentally (chaotic structures with adjustable point contacts, including tunable tunnel barriers, can by now be fabricated\cite{KMR-1997}). In this Letter, a first {\it systematic} foray is made into a largely unexplored territory of non-ideal couplings. In doing so, we choose (for the sake of simplicity) to lift a point-contact-ballisticity only for the left lead which is assumed to support $n_{\rm L}$ propagating modes characterised by a set of tunnel probabilities $\hat{{\bm \Gamma}}_{\rm L}=(\Gamma_1,\dots,\Gamma_{n_{\rm L}})$; the right lead, supporting $n_{\rm R} \ge n_{\rm L}$ open channels\cite{R-01}, is kept ideal so that $\hat{{\bm \Gamma}}_{\rm R}=(\Gamma_{n_{\rm L}+1},\dots,\Gamma_{n_{\rm R}})=\openone_{n_{\rm R}-n_{\rm L}}$. Assuming that the time-reversal symmetry is broken ($\beta=2$), we shall show that the joint probability density function $P_{(\hat{{\bm \gamma}}_{\rm L}|\,{\bm 0})}({\bm R})$ of reflection eigenvalues $\{R_j=1-T_j \}$ equals\cite{R-01a}
\begin{widetext}
\begin{equation}\label{jpdf-1}
   P_{(\hat{{\bm \gamma}}_{\rm L}|\,{\bm 0})}(R_1,\dots,R_{n_{\rm L}})  = c_{n_{\rm L},n_{\rm R}} \,
   \frac{
   {\rm det}{}^{N} (\openone_{n_{\rm L}} - \hat{{\bm \gamma}}_{\rm L}^2)
    }{\Delta_{n_{\rm L}}(\hat{{\bm \gamma}}_{\rm L}^2)}
    \, \Delta_{n_{\rm L}} ({\bm R}) \,
            {\rm det}_{(j,k)\in (1, n_{\rm L})} \Big[
                {}_2 F_1 (n_{\rm R}+1,n_{\rm R}+1; 1;\, \gamma_j^2 R_k)
            \Big]\prod_{j=1}^{n_{\rm L}} (1-R_j)^{\nu}.
\end{equation}
\end{widetext}
Here, $\hat{{\bm \gamma}}_{\rm L}^2 = \openone_{n_{\rm L}}- \hat{{\bm \Gamma}}_{\rm L}$ is a set of $n_{\rm L}$ coupling parameters characterising non-ideality of the left lead in terms of associated tunnel probabilities, $N=n_{\rm L}+n_{\rm R}$ is the total number of open channels in both leads, $c_{n_{\rm L},n_{\rm R}}$ is the inverse normalisation constant, \begin{eqnarray}
    c_{n_{\rm L},n_{\rm R}} = \frac{(n_{\rm L}+n_{\rm R})!}{n_{\rm L}! \, n_{\rm R}!}
    \prod_{j=1}^{n_{\rm L}} \frac{(n_{\rm R}!)^2}{(n_{\rm R}+j)! \, (n_{\rm R}-j)!},
\end{eqnarray}
whilst ${}_p F_q$ is the Gauss hypergeometric function. The (biorthogonal \cite{DF-2008}) ensemble of reflection eigenvalues Eq.~(\ref{jpdf-1}) is our {\it first main result}\cite{Remark0}. Before outlining its derivation, let us discuss the implications of Eq.~(\ref{jpdf-1}) for a nonperturbative statistical description of {\it both} spectral and transport observables in quantum chaotic cavities.

{\it Statistics of reflection eigenvalues.}---The first immediate consequence of Eq.~(\ref{jpdf-1}) is the determinant structure of the $p$-point correlation function of reflection eigenvalues:
\begin{equation}\label{cf}
    \rho_{(n_{\rm L},n_{\rm R})}(R_1,\dots,R_p) = {\rm det}_{(j,k)\in(1,p)}\left[ K_{(n_{\rm L},n_{\rm R})}(R_j,R_k%; \hat{{\bm \gamma}}_{\rm L}^2
    ) \right].
\end{equation}
Defined in a standard manner\cite{M-2004}, it is entirely determined by the two-point scalar kernel $K_{(n_{\rm L},n_{\rm R})}(R,R^\prime)$, that can straightforwardly be calculated \cite{VK-2011} by applying the ideas exposed in Ref.\cite{DF-2008}. In terms of the ``moment function''
\begin{eqnarray}\label{mk-nu}
    M_{k}^{(\nu)}(\gamma^2) =  \sum_{\ell=0}^\nu (-1)^\ell \left(\atop{\nu}\ell\right) \, M_{k+\ell}^{(0)}(\gamma^2),
\end{eqnarray}
where $M_{k}^{(0)}(\gamma^2) = k^{-1} \, {}_3 F_2 (n_{\rm R}+1,n_{\rm R}+1,k; 1,k+1;\gamma^2)$, the scalar kernel is given by a finite sum:
\begin{widetext}
\begin{eqnarray}\label{kernel}
    K_{(n_{\rm L},n_{\rm R})}(R,R^\prime) &=&
    n_{\rm L}! \,c_{n_{\rm L},n_{\rm R}}  \, \frac{{\rm det}^{n_{\rm L}+n_{\rm R}} (\openone_{n_{\rm L}} - \hat{{\bm \gamma}}_{\rm L}^2)}{\Delta_{n_{\rm L}}(\hat{{\bm \gamma}}_{\rm L}^2)}
    \, [(1-R)(1-R^\prime)]^{\nu/2}\\
    &\times&
    \, \sum_{j=1}^{n_{\rm L}} {}_2 F_1 (n_{\rm R}+1,n_{\rm R}+1,1;\gamma_j^2 R)
    \, {\rm det} \left[
    [M_{k}^{(\nu)}(\gamma_\ell^2)]_{\ell=1,\cdots,j-1}; \, (R^\prime)^{k-1}; \, [M_{k}^{(\nu)}(\gamma_\ell^2)]_{\ell=j+1,\cdots,n_{\rm L}}
    \right].\nonumber
\end{eqnarray}
\end{widetext}
Here, $k \in (1, n_{\rm L})$ counts the rows of an $n_{\rm L}\times n_{\rm L}$ matrix under the sign of determinant. Equations (\ref{cf}) and (\ref{kernel}) represent our {\it second main result}.

\begin{figure}[t]
\includegraphics[scale=0.51]{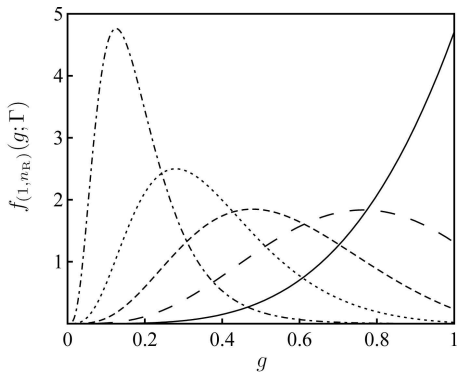}
\caption{Probability density function $f_{(1,n_{\rm R})}(g;\Gamma)$ for Landauer conductance plotted for $n_{\rm R}=5$ and various tunnel probabilities: $\Gamma=0.99$ (solid line), $\Gamma=0.8$ (long-dashed line), $\Gamma=0.6$ (dashed line), $\Gamma=0.4$ (dotted line), and $\Gamma=0.2$ (dot-dashed line). For the ideal point contact ($\Gamma=1$), the curve is a monotonous function of $g$ reaching its maximum at $g_*=1$. Decrease of the tunnel probability leads to development of a well-pronounced maximum. In the large-$n_{\rm R}$ limit, it is positioned at $g_*=\Gamma$, see Eq.~(\ref{g*}).}\label{Fig-1}
\end{figure}

{\it Distribution of Landauer conductance}.---Although the central result of this Letter, Eq.~(\ref{jpdf-1}), allows us to address the problem of conductance fluctuations in full generality, the most explicit formulae can be obtained for the illustrative example of an asymmetric cavity whose left (non-ideal) lead supports a single propagating mode ($n_{\rm L}=1$). For such a setup, a probability density $f_{(1,n_{\rm R})}(g;\Gamma)$ of the Landauer conductance is proportional to the mean density $K_{(1,n_{\rm R})}(R,R)$ of reflection eigenvalues taken at $R=1-g$. Materialising this observation with the help of Eq.~(\ref{kernel}), we derive \cite{R-particular}:
\begin{eqnarray}\label{cond-d}
    f_{(1,n_{\rm R})}(g;\Gamma) = f_{(1,n_{\rm R})}(g;1)\, \Gamma^{n_{\rm R}+1}\qquad\qquad\qquad\qquad\nonumber \\
    \times \,
    {}_2 F_1 (n_{\rm R}+1,n_{\rm R}+1; 1;\, (1-\Gamma)(1-g)). \quad
\end{eqnarray}
Here, $\Gamma$ is the tunnel probability of the left point contact, whilst
$f_{(1,n_{\rm R})}(g;1) = n_{\rm R} g^{n_{\rm R}-1}$
describes the conductance density when the left point contact is ballistic.

The probability density of Landauer conductance Eq.~(\ref{cond-d}) shows an unusually rich behavior (see Fig.~\ref{Fig-1}). First, it exhibits a pronounced maximum whose position $g_*$, for a {\it generic value} of the tunnel probability $\Gamma$, depends on the number ($n_{\rm R}$) of propagating modes in the ideal lead. Second, numerical analysis of Eq.~(\ref{cond-d}) reveals existence of a ``critical'' value ($\Gamma_0$) of the tunnel probability: for $\Gamma < \Gamma_0$, increase of $n_{\rm R}$ makes the maximum position move from {\it left to right} until it approaches its saturated location $g_*=\Gamma$; on the contrary, for $\Gamma > \Gamma_0$, as $n_{\rm R}$ increases, position of the maximum moves in the opposite direction eventually reaching $g_*=\Gamma$.

\begin{figure}[b]
\includegraphics[scale=0.52]{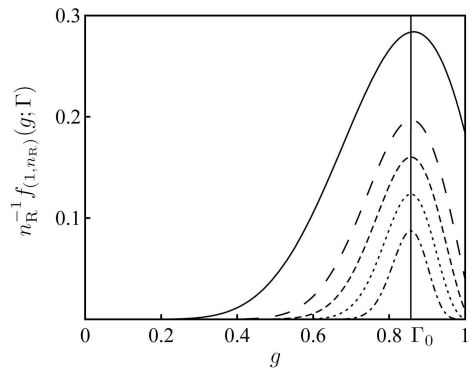}
\quad
\caption{Unnormalised conductance ``distribution'' $n_{\rm R}^{-1} f_{(1,n_{\rm R})}(g;\Gamma_0)$ plotted for the ``critical'' tunnel probability $\Gamma_0=6/7$ and various numbers of propagating modes $n_{\rm R}$ in the ideal lead: $n_{\rm R}=10$ (solid line), $n_{\rm R}=20$ (long-dashed line), $n_{\rm R}=30$ (dashed line), $n_{\rm R}=50$ (dotted line), and $n_{\rm R}=100$ (dot-dashed line). Position of the maximum is almost ``frozen'', depending very weakly on $n_{\rm R}$ [see Eq.~(\ref{g*})].} \label{Fig-2}
\end{figure}

To describe this effect analytically, one has to seek an explicit functional form of $g_*(\Gamma, n_{\rm R})$ for arbitrary $\Gamma$ and $n_{\rm R}$, which appears to be an impossible task. However, some progress can be made in the large-$n_{\rm R}$ limit, when an $1/n_{\rm R}$ expansion can be developed. A somewhat cumbersome calculation \cite{VK-2011} based on the asymptotic analysis of the hypergeometric function in Eq.~(\ref{cond-d}) brings out the remarkable formula
\begin{eqnarray}\label{g*}
    g_*(\Gamma, n_{\rm R}) = \Gamma \left[
        1 + \frac{7}{2 n_{\rm R}} \left(
            \Gamma - \frac{6}{7}
        \right) + O(n_{\rm R}^{-2})
    \right],
\end{eqnarray}
suggesting that the ``critical'' value $\Gamma_0$ of the tunnel probability equals $\Gamma_0=6/7$. This prediction is unequivocally supported by numerics based on the exact Eq.~(\ref{cond-d}), see Fig.~\ref{Fig-2}. We believe that experimental testing of the ``$6/7$'' effect may be feasible within the current limits of nanotechnology \cite{E-2001}.

Finally, we mention that a calculation of $f_{(n_{\rm L},n_{\rm R})}(g;\Gamma)$ becomes increasingly complicated for $n_{\rm L}>1$. This difficulty, however, can be circumvented by focussing on the moment/cumulant generating function \cite{OK-2008} of the Landauer conductance, that can be related (under certain assumptions) to solutions of the two-dimensional Toda Lattice equation \cite{VK-2011}.

{\it Sketch of the derivation.}---Having discussed a few (out of potentially many) implications of the joint probability density of reflection eigenvalues in chaotic cavities probed via both ballistic and tunnel point contacts [Eq.~(\ref{jpdf-1})], let us outline its derivation. The $\beta=2$ Poisson kernel Eq.~(\ref{pk}) with the average scattering matrix $\bar{{\bm {\mathcal S}}}$ set to\cite{R-02}
%\begin{eqnarray}\label{asm}
$\bar{{\bm {\mathcal S}}}= {\rm diag}(\hat{{\bm \gamma}}_{\rm L}, 0 \times \openone_{n_{\rm R}})$
%\end{eqnarray}
and a polar-decomposed \cite{H-1963,F-2006} unitary scattering matrix ${\bm {\mathcal S}}$,
\begin{eqnarray}\label{pd-1}
    {\bm {\mathcal S}} = \left(
                           \begin{array}{cc}
                             {\bm u}_1 & 0 \\
                             0 & {\bm v}_1 \\
                           \end{array}
                         \right)
                             \hat{{\mathcal L}}({\bm \lambda})
                         \left(
                           \begin{array}{cc}
                             {\bm u}_2 & 0 \\
                             0 & {\bm v}_2 \\
                           \end{array}
                         \right)
\end{eqnarray}
is our starting point. Here,
\begin{eqnarray}\label{pd-2}
        \hat{{\mathcal L}}({\bm \lambda}) = \left(
                           \begin{array}{cc}
                             \sqrt{\openone_{n_{\rm L}} - {\bm \lambda}{\bm \lambda}^{\rm T}} & i{\bm \lambda} \\
                              i{\bm \lambda}^{\rm T} & \sqrt{\openone_{n_{\rm R}} - {\bm \lambda}^{\rm T}{\bm \lambda}} \\
                           \end{array}
                         \right),
\end{eqnarray}
the matrix ${\bm \lambda}$ is an $n_{\rm L} \times n_{\rm R}$ rectangular diagonal matrix such that ${\bm \lambda}{\bm \lambda}^{\rm T} = {\rm diag}(T_1,\dots, T_{n_{\rm L}})$ if $n_{\rm L} \le n_{\rm R}$, and ${\bm \lambda}{\bm \lambda}^{\rm T} = {\rm diag}(T_1,\dots, T_{n_{\rm R}}; \, 0 \times \openone_{n_{\rm L}-n_{\rm R}})$ otherwise; the matrices ${\bm u}_j$ and ${\bm v}_j$ are unitary matrices of the size $n_{\rm L}\times n_{\rm L}$ and $n_{\rm R}\times n_{\rm R}$, respectively. Restricting ourselves to a structurally more transparent case \cite{R-01} $n_{\rm L} \le n_{\rm R}$, we notice that the polar decomposition induces the relation
\begin{equation}\label{jac}
    d\mu({\bm {\mathcal S}}) = P_{0}(\openone_{n_{\rm L}} - {\bm R}) \prod_{j=1}^{n_{\rm L}} dR_j  \prod_{\alpha=1}^2 d\mu({\bm u}_\alpha)\, d\mu({\bm v}_\alpha),
\end{equation}
where $P_{0}(\openone_{n_{\rm L}} - {\bm R})=P_{0}({\bm T})$ is the joint probability density function of transmission eigenvalues at $\beta=2$ in case of ideal leads [Eq.~(\ref{jpdf-1})], and $d\mu$ is the invariant Haar measure on the unitary group.

Substituting Eqs.~(\ref{pd-1}) and (\ref{pd-2}) into Eq.~(\ref{pk}) taken at $\beta=2$, and considering the elementary volumes identity Eq.~(\ref{jac}), we conclude that the j.p.d.f. of reflection eigenvalues in the non-ideal case admits the representation
\begin{eqnarray}\label{jpdf-1a}
    P_{(\hat{{\bm \gamma}}_{\rm L}|\,{\bm 0})}({\bm R}) &\propto& {\rm det}^N(
        \openone_{n_{\rm L}} -  \hat{{\bm \gamma}}_{\rm L}^2)\, P_{0} (\openone_{n_{\rm L}} - {\bm R})\,
        \nonumber\\
        &\times&
         \int_{{\rm U}(n_{\rm L})} d\mu({\bm U}) \int_{{\rm U}(n_{\rm L})} d\mu({\bm V})\,  \nonumber \\
    &\times&\det{}^{-{N}}(\openone_{n_{\rm L}}- \hat{{\bm \gamma}}_{\rm L}\bm{U} \hat{\bm \varrho} \bm{V}^\dagger) \nonumber\\
    &\times&\, \det{}^{-N}(\openone_{n_{\rm L}}- \bm{V} \hat{{\bm \varrho}}
    \,\bm{U}^\dagger \hat{{\bm \gamma}}_{\rm L}),
\end{eqnarray}
where the notation $\hat{\bm \varrho}$ stands for $\hat{\bm \varrho}=\sqrt{\openone_{n_{\rm L}} - {\bm \lambda}{\bm \lambda}^{\rm T}}={\rm diag}(R_1^{1/2},\dots, R_{n_{\rm L}}^{1/2})$. Notice, that for any finite $\hat{{\bm \gamma}}_{\rm L}$, the ${\rm U}(n_{\rm L})\times {\rm U}(n_{\rm L})$ group integrals in Eq.~(\ref{jpdf-1a}) effectively modify the interaction between reflection eigenvalues, which is no longer logarithmic [see Eq.~(\ref{PnT})].

The group integrals in Eq.~(\ref{jpdf-1a}) can be evaluated by employing the technique of Schur functions\cite{M-1995} and the theory of hypergeometric functions of matrix argument\cite{M-2005,GR-1989}. (An alternative derivation, based on the theory of $\tau$ functions of matrix argument, was reported in Ref. \cite{O-2004}.) Leaving details of our calculation for a separate publication \cite{VK-2011}, we state the final result:
\begin{equation} \label{final-dgi}
    \frac{ {\rm det}_{(j,k)\in (1, n_{\rm L})} \Big[
                {}_2 F_1 (n_{\rm R}+1,n_{\rm R}+1; 1;\, \gamma_j^2 R_k)
            \Big]}{\Delta_{n_{\rm L}}(\hat{{\bm \gamma}}_{\rm L}^2)\, \Delta_{n_{\rm L}}({\bm R})}.
\end{equation}
Combining the last two equations together, we reproduce the joint probability density function of reflection eigenvalues announced in Eq.~(\ref{jpdf-1}).

{\it Summary.}---In this Letter, we have outlined an RMT approach to the problem of universal quantum transport in chaotic cavities probed through both ballistic and tunnel point contacts. While our central result Eq.~(\ref{jpdf-1}) marks quite a progress in equipping the field with nonperturbative calculational tools, certainly more efforts are required to bring the theory to its culminating point: (i) relaxing a point-contact-ballisticity for the second lead, (ii) extending the formalism to other Dyson-Altland-Zirnbauer symmetry classes \cite{AZ-1997,B-2009}, and (iii) studying integrable aspects of the theory, much in line with Ref. \cite{OK-2008}, is just a partial list of related challenging problems whose solution is very much called for.

This work was supported by the Israel Science Foundation through the grant No 414/08. 

{\it Note added in proof}.---Recently, we have learned about
the paper by Y.~V.~Fyodorov \cite{F-2003} who studied a ''probability
of no-return'' in quantum chaotic and disordered systems. In a certain limit, this probability can be reinterpreted
as the Landauer conductance distribution $f_{(1,n_{\rm R})}(g;\Gamma)$
given by Eq.~(\ref{cond-d}) of this Letter. We have explicitly verified
that both results are equivalent to each other.


\begin{references}
\bibitem{R-01a} To simplify the notation, we have dropped the superscript $(\beta=2)$. The subscript $(\hat{{\bm \gamma}}_{\rm L}|\,{\bm 0})$ is used to explicitly indicate a non-ideal (ideal) character of the left (right) lead.
\bibitem{Remark0} In the limit $\hat{{\bm \gamma}}_{\rm L} \rightarrow {\bm 0}$, the ratio of the ${}_2 F_1$-determinant to $\Delta_{n_{\rm L}}(\hat{{\bm \gamma}}_{\rm L}^2)$ becomes proportional to the Vandermonde determinant $\Delta_{n_{\rm L}}({\bm R})$ thus reproducing the $\beta=2$ variant of Eq.~(\ref{PnT}) derived for the case of ideal leads.
\end{references}
\end{document}